\documentclass[envcountsame]{llncs}

\usepackage{graphicx}

\title{Human machine epistemology survey}
\author{R\'{e}mi Nazin\inst{1} \inst{2} \thanks{Supported for a PhD by the Direction G\'{e}n\'{e}rale de l'Armement.} \and Didier Fass \inst{3}\inst{2}
}
\institute{PErSEUs (EA 7312) - Universit\'{e} de Lorraine 
\and LORIA -  MOSEL \email{remi.nazin@loria.fr}, \email{didier.fass@loria.fr} \and ICN Business School Nancy}

\begin{document}
\maketitle

\begin{abstract}
Pluridisciplinar convergence is a  major problem that had emerged with human-artefact Systems and so-called "Augmented Humanity" as academical fields and even more as technical fields. Problems come mainly from the juxtaposition of two very different types of system, a biological one and an artificial one. Thus, conceiving and designing the multiple couplings between them has become a major difficulty. Some came with reductionnist solutions to answer these problems but since we know that a biological system and a technical system are different, this approach is limited from its beginning. 

Using a specifically designed questionnaire and statistical analysis we determined how specialists (medical practitioners, ergonomists and engineers) in the domain conceive themselves what is a human-artifact System and how they relate to existent traditions and we showed that some of them relate to the integrativist views.
\end{abstract}

\section{Introduction}
\subsection{The integrative way of looking at things}
Designing human-artefact Systems within the current technological context impose to adress safety and reliability issues at the same time\footnote{Particularly in the medical domain.}.  

Some theoretical apparatus are currently in use to support the design of human-artefact Systems but either they are ``incomplete\footnote{They can answer only a fraction of the problems.}'' of they adopt some form of reductionnism. This situation tends to lead researchers and engineers to build their own composite theories on demand thus being exposed to underlying contradiction which represents a critical safety issue. 

Therefore, a unified but comprehensive way of conceiving what is a human-artefact System is necessary. This way is that of integration as as intellectual approach which explains the functionning of a given system by those of its components and their organization. As the philosopher once said this could be regarded as a\emph{"new name for some old ways of thinking"\footnote{William James as a subtitle for his \emph{Pragmatism}.}} but the important word in the definition is \textbf{system}. What we mean by \textbf{system} is linked to the general system theory (GST)\cite{bertalanffy2012} and designate a complex of objects whose interactions give new properties to the whole. 

Apart from its systemical dimension, the integrative way of looking at human-artefact Systems requires to be a truly trandisciplinary paradigm. That means that it should allow us to understand the systems from their physical to their social and logical dimensions via their biological or artificial dimension, without getting out of our framework. 

\subsection{The generic system}
In his article(\cite{fass2012augmented}), Fass builded a new definition of a generic and isomorphic framework for describing systems. This system is defined by its dimensions of requirement (shape, dynamics and elements) and of specification (architecture, evolution and behaviour). The interesting point about specifications is their ability to characterize the functional identity of the modeled system including its behaviours and evolutions. It should be noted that this definition of a system is perfectly coherent with the theory of Bertalanffy (\cite{bertalanffy2012}) and share with it the fact that the system has no $\tau \acute{\varepsilon} \lambda o \varsigma $\footnote{$\tau \acute{\varepsilon} \lambda o \varsigma $ is the ancient greek for ``completion'' and is commonly used to describe the target state of a system as a metaphysical attribute.} but is equifinal\footnote{An equifinal system tend to a characterstical state from several initial states and various routes.}.

For more considerations upon the epistemological implications of this redefinition the reader can consult the Master thesis of the main author\footnote{\cite{nazin2014}}.

\begin{figure}
\includegraphics{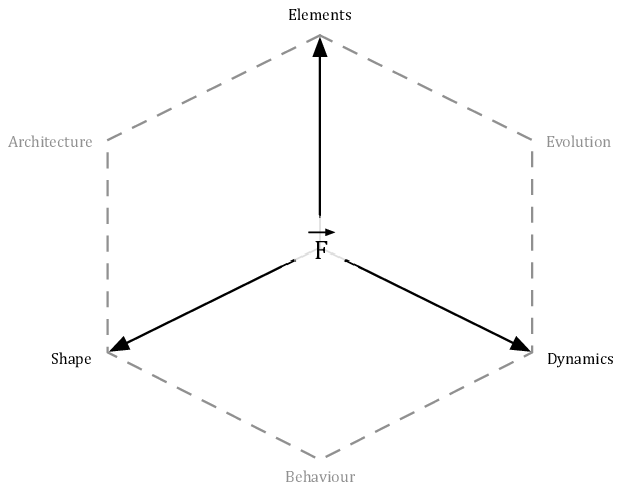}
\caption[]{The generic system and its dimensions.

\smallskip

Plain lines represents requirements, dotted lines represents specifications. We can see that specifications are produced by the intersection of requirements and that the function is obtained from the global behaviour of the system.}
\label{cube}
\end{figure}
\subsection{The link between theory and practice}
In order to demonstrate that there is some space available for our framework in the scientifical panorama, we had to adress a difficult problem. Medical practitioners, engineers and ergonomists have indeed little or no knowledge in epistemology because it is not part of their technical education. Being uneducated in a discipline does not mean having no idea about it, it does rather mean having ideas which are not necessarily organized in a consistant way. 

There is an obvious link between theory and practice in the sense that some theoretical ideas that are not directly from one's field of expertise\footnote{e.g. "consciousness is alike of a computer program" for a medical doctor.} can impact his practice. These theoretical ideas are entangled in a more or less coherent way into an underlying theoretical background. This background is not spontaneously created but is built during the education, experience and scientific general culture. 

It is therefore impossible to ask our population to which tradition they find themselves the most affiliated. This situation imposes us then to probe their underlying theoretical background as a scientific ideology. 

We first established a list of the main traditions used to conceive human artefact Systems. This list is constituted by behaviorism, cognitivism, connexionnism and cybernetics. For comparison, we added a brand new ``tradition'' called \emph{integrativism} which reflects our point of vue upon the nature of human hachine Systems.

It should be duly noted that, although the structure is inherited from our definition of a generic system, the content of the list is not arbitrary but comes from our bibliographical researches.

We then listed the respective theoretical positions upon the major items of the domain if possible\footnote{Meaning there is a position.}. 

From the bibliography, we could build a table to compare what we call \emph{canonical traditions}. A canonical tradition reflects the core principles of a scientific movement about a given subject,  the actual positions taken by an individidual may not be exactly the same as the canonical tradition which he is related because of the existence of different currents inside a given tradition. This is why we focus on the core principles which are the one that allows to distinguish some groups inside our population.

\begin{table}
\caption[]{Canonical traditions on the nature of human-machine systems (part 1)}
\begin{tabular}{ll|lll}

\hline\noalign{\smallskip}

      &   & Behaviourism & Cognitivism & Connexionism \\ 
\hline

    Individual &   &   &   &     \\ 

  & \shortstack[l]{Body\\$ $\\$ $\\$ $} & \shortstack[l]{Sensory organs \\+ Effector organs \\+ Nervous system } & \shortstack[l]{Biological computer \\+ Sensory Organs  }& \shortstack[l]{Neural network \\+ Body }\\ 

  &   &   &   &   \\ 

  & \shortstack[l]{Mind\\ $ $\\ $ $\\ $ $} & \shortstack[l]{Subvocal language \\+ Memory} & \shortstack[l]{Centralized symbolic\\ information processing \\+ Memory }& \shortstack[l]{Emergent patterns \\in the network }\\ 

  &   &   &   &   \\   &   &   &   &   \\ 

Architecture &  & Mechanistic & Centralized modules  & Network\\ &   &   &   &\\ 

  & Constitution & Conditioning & / & Reinforcement \\ &   &   &   & \\ 
 
  & Element & Spatial sub-unit & Module & \shortstack[l]{Connex part \\of the network }\\ 

  &   &   &   &   \\ &   &   &   &   \\ 

Behaviour &   & \shortstack[l]{Articulated response\\ to stimulus} & Problem solving & Computation \\ &   &   &   &   \\ 

  & \shortstack[l]{Cognition\\ $ $\\ $ $} & \shortstack[l]{Articulated response\\ to stimulus } & \shortstack[l]{Rule based \\symbols manipulation} & \shortstack[l]{Adequate pattern\\ activation }\\ &   &   &   &  \\ 

  & Perception & Stimulus recognition  & Information acquisition & Information acquisition \\ &   &   &   &     \\ 
  
  & \shortstack[l]{Action\\ $ $\\ $ $ }& \shortstack[l]{Articulated response\\ to stimulus } & Information treatment result & Computational result \\ 

  &   &   &   &   \\ &   &   &   &   \\ 

Evolution &   & Mechanistic processes  & Algorithmic sequences & Activation sequences \\ &   &   &   &  \\ 

  & \shortstack[l]{Interaction\\ $ $\\ $ $} & \shortstack[l]{Action and reaction\\ in the environment } & Data treatment & Input/Output  \\ &   &   &   &   \\ 

  & Memory & Faculty of the mind  & Faculty of the mind & Patterns of activation \\ 
  &   &   &   &   \\ &   &   &   &    \\ 

Function &   & None  & Related to a module& / \\ 

  &   &   &   &    \\ &   &   &   &    \\ 

Communication &   & Inter-individual   & Information Theory & Physico-chemical signal \\ 

  &   &   &   &    \\ &   &   &   &    \\ 

Environnment &   & Stimulus source  & \shortstack[l]{Source of information\\ and problems} & \shortstack[l]{Source of information\\ and problems} \\ 

\end{tabular}
\end{table}

\begin{table}
\caption[]{Canonical traditions on the nature of human-machine systems (part 2)}
\begin{tabular}{ll|ll}

\hline\noalign{\smallskip}

      &   &  Cybernetics & Integrativism \\ 
\hline

    Individual &   &   &   \\ 

  & Body & Feedback regulated systems & \shortstack[l]{Self-associated \\metastable systems \cite{chauvet2006}}\\ 
  &   &    &   \\

  & \shortstack[l]{Mind\\ $ $\\ $ $} & \shortstack[l]{Decentralized information processing \\+ Memory} & \shortstack[l]{Consciousness \\(In a Jamesian way)\cite{edelman1992}} \\ 
 
   &   &    &   \\ 
  &   &    &   \\

Architecture &  & Feedback loops network & Biological \\ 
  & Constitution & / & Stabilizating Self Association \cite{chauvet1993hierarchical}\\
   &   &    &   \\ 

  & Element & Feedback loop  & Fonctionnal sub-unit \\ 

  &   &    &   \\ 
  &   &    &   \\

Behaviour &  &  Return to equilibrium &   \\ 

  & \shortstack[l]{Cognition \\ $ $\\ $ $} & / & \shortstack[l]{Categorization by \\thalamo-cortical reentry \cite{edelman2004}}\\  &   &    &   \\ 

  & Perception & Environmental disturbance & Integration of sensitive data\cite{gibson2011} \\  &   &    &   \\

  & Action & Recherche d'équilibre &  Domain of stability and viability  \\ 

  &   &    &   \\ 
  &   &    &   \\

Evolution &   & Return to equilibrium & Sub-system interactions \\ 

  & Interaction &  & Transductive coupling  \cite{chauvet1993hierarchicalc}\cite{chauvet1993hierarchicalb}\\   &   &    &   \\

  & Memory & / & Recategorization \cite{edelman1992}\\ 

  &   &    &   \\ 
  &   &    &   \\

Function &   & Regulation & Resulting from the activity of the system\\ 
  &   &    &   \\ 
  &   &    &   \\

Communication &   & Information Theory & Functional interaction \cite{chauvet2006}\\ 
  &   &    &   \\ 
  &   &    &   \\

Environnment &  & Source of perturbations & Hypersystem \\ 

\end{tabular}
\end{table}

\section{Hypothesis}
The survey was designed to test the following hypothesis in a qualitative manner. 
\begin{enumerate}
\item[(h0)]{The population is divisible into different coherent groups in terms of their sets of answers.}
\item[(h1)]{If (h0) is verified, at least one coherent group can be related to a canonical tradition.}
\item[(h2)]{If (h1) is verified, there is no necessary link between the field of expertise and the membership to a coherent group.}
\item[(h3)]{If (h2) is verified, there is a coherent group related to integrativism.}
\end{enumerate}

\section{Materials used}
\begin{enumerate}
\item[-]{The Inria enquiry plateform runs LimeSurvey 2.05+\footnote{https://www.limesurvey.org}.}
\item[-]{Statistical treatments of the answers were made with R 3.0.2\footnote{http://www.R-project.org} on OS X 10.9.}
\item[-]{Correlation graphs were made with Gephi 0.8.2\footnote{https://gephi.github.io} on OS X 10.9.}
\end{enumerate}

\section{Method}
From our hypothesis, we see that the purpose of the study was to show the existence of some sets of scientifical beliefs which are not directly provided by a domain of expertise and that these sets are shared by people form very different domains. To do this, a qualitative analysis must be conducted.
\subsection{Survey}
To examine our hypotheses, we selected some items in our grid of canonical traditions and we transformed them into affirmative propositions. To eliminate every potential bias, we randomized the initial order of the propositions.  What we wanted to measure was the degree of agreement of our subjects upon a series of these propositions. For reasons of convenience, we choose to limit the length of the questionnaire to 25, keeping the completing-time under 20 minutes. For measuring the agreement, we used the typical format of a five item Lickert scale (\cite{likert1932})\footnote{Strong disagreement, simple disagreement, without idea, agreement, strong agreement.}. 

We completed the questionnaire with some demographic items such as sex, age and professional occupation\footnote{This item has been used as an open question in regard of the means of transmission of the questionnaire. We wanted to touch a maximal variety of subjects and the use of predetermined categories could have exluded a part of our general population.}.  These questions allowed us to refine our results and to test (h4). Using the LimeSurvey plateform, we guaranteed the anonimity of our subjects by not registering their ip adress and not using cookies. The transcription of the questionnaire is available in figure 2.
 
The questionnaire was available to all public trough the dedicated Inria plateform\footnote{https://sondages.inria.fr} during a 15 days period. The plateform uses a encrypted connection and thus reinforce the anonimity of our subjects.  In addition to the common traffic of the plateform, we broadcasted the questionnaire through selected diffusion lists and communities :
\begin{enumerate}
	\item[-]{[tousloria] features all the people of the LORIA.}
	\item[-]{[ergoihm] features a great variety of ergonomicians and specialists in HCI.}
	\item[-]{[info-ic] feature the community of researchers in knowledge engineering.}
	\item[-]{La Soci\'{e}t\'{e} de r\'{e}animation de Langue Fran\c{c}aise.}
\end{enumerate}

\section{Statistical Analysis}
\subsection{Principal Component Analysis}
Principal Component Analysis (PCA) is a factorial analysis method. Its main strength is its ability to describe an individual-variables matrix without any statistical hypothesis. As a factorial analysis method, it is not impacted by the size of the population either.

We applied this method to our groups of population by testing the $\theta _0$ hypothesis that the absence of principal component significantly distinguishable in a population is a sign of the existence of some groups inside it. This hypothesis is opposed to $\theta _1$ that the existence of a significant principal component in a group allows to define a group of answers greatly influenced by some determined variables and by then that the group is homogenous.
\subsection{Hierarchical clustering and classification}
This method allows to divide a group into several more homogenous groups in regard of the variation of the eigenvalue. It produces a hierarchical tree of clusters which can be analyzed by other methods described below. 
\subsection{Graphs of correlation}
As all our variables corresponding to answers are of the same type, it is possible to apply a $\chi ^2$ test\footnote{With a p-value of 5\%} for each couple of them. We can thus build a correlation matrix and a correlation graph. The correlation graph represents the network of reciprocal influences of the answers in a given group. Each node represents an answer and each edge represents a correlation between two nodes\footnote{The qualitative dimension of the correlation is not of interest here in so far as we want to represent the absolute influence of each variable on the others.}.

This tool is interesting because it is more easily readable than a matrix and has graph properties such as connectivity. The more connected is the correlation graph, the more consistent is the group of answers. Then a graph with no connectivity at all represent the case of random answering and a graph fully connected represent the case of a group who has answered all the questions with the same answer. 

\begin{figure}
	\label{survey}

\begin{enumerate}	

	\item What is our age ?
	\item What is your sex ?
	\item What is your profession ?
	\item[]{} 
	\item A machine may be compared to an organ.
	\item Perceptual information treatment is the main function of the brain.
	\item Behaviour is solely a reaction to exterior stimulus.
	\item It is possible to predict the behaviour of an individual regardless of its body.
	\item Reasoning could be considered as the result of a logical calculus.
	\item It is possible to study an individual disregarding its interactions with the environnment.
	\item A living cell may be compared to a machine.
	\item Mind est reducible to symbol manipulation.
	\item A model which describes the augmented human must necessary be of biological inspiration.
	\item An organ can be compared to a machine.
	\item Behaviour cans be described as a problem resolution situation. 
	\item Environnement can be viewed solely as a stimuli source.
	\item It is possible to understand the behaviour of an individual regardless of its nervous system.
	\item Brain can be described as a computer.
	\item We can describe the fonctionnement of an organism by a group of functions.
	\item Augmented human is limited to enhance its natural functions.
	\item Perception is a passive acquisition of information.
	\item Capacity enhancement is the begining of a robotization of humanity.
	\item Restoring a damaged biological function is enhancement.
	\item Logic can account for the whole mind.
	\item Sensorial information is tranformed during perception.
	\item Information from the environnment can be regarded as a stimulus.
	\item It is possible de reduce the beahviour of a neuron to that of a logical operator.
	\item Enhanced humanity is linked to new functions.
	\item Capacity enhancement impose deep modifications in the subject.

	\end{enumerate}
	
	\caption[]{Questionnaire used in this research}
	
	\end{figure}

\section{Results}
\subsection{Description of the population}
After closing of the questionnaire, we had 150 complete sets of answers divisible in three categories (medical practitioners : 36 , ergonomists : 59 , others : 55). This division of the population doesn't impact the study in so far as it is used in the very end of it in order to obtain the final results displayed in figure \ref{results}.

Concerning our medical population, its size is of 36 which allowed us to consider our goals as achieved and it validate the population since the size of the general population is sufficient. The ``other'' category means to represent the non expert part of the population and its number is explained by the large opening of the survey via the plateform.

In the global population, we find two men for a woman and the majority of the subjects has an age between 18 and 55 years with a spike between 25 and 30 years. In the light of socio-demographic researches concerning the scientific population in France, this enables us to qualitatively validate the representativity of the general population.

\begin{figure}
\includegraphics{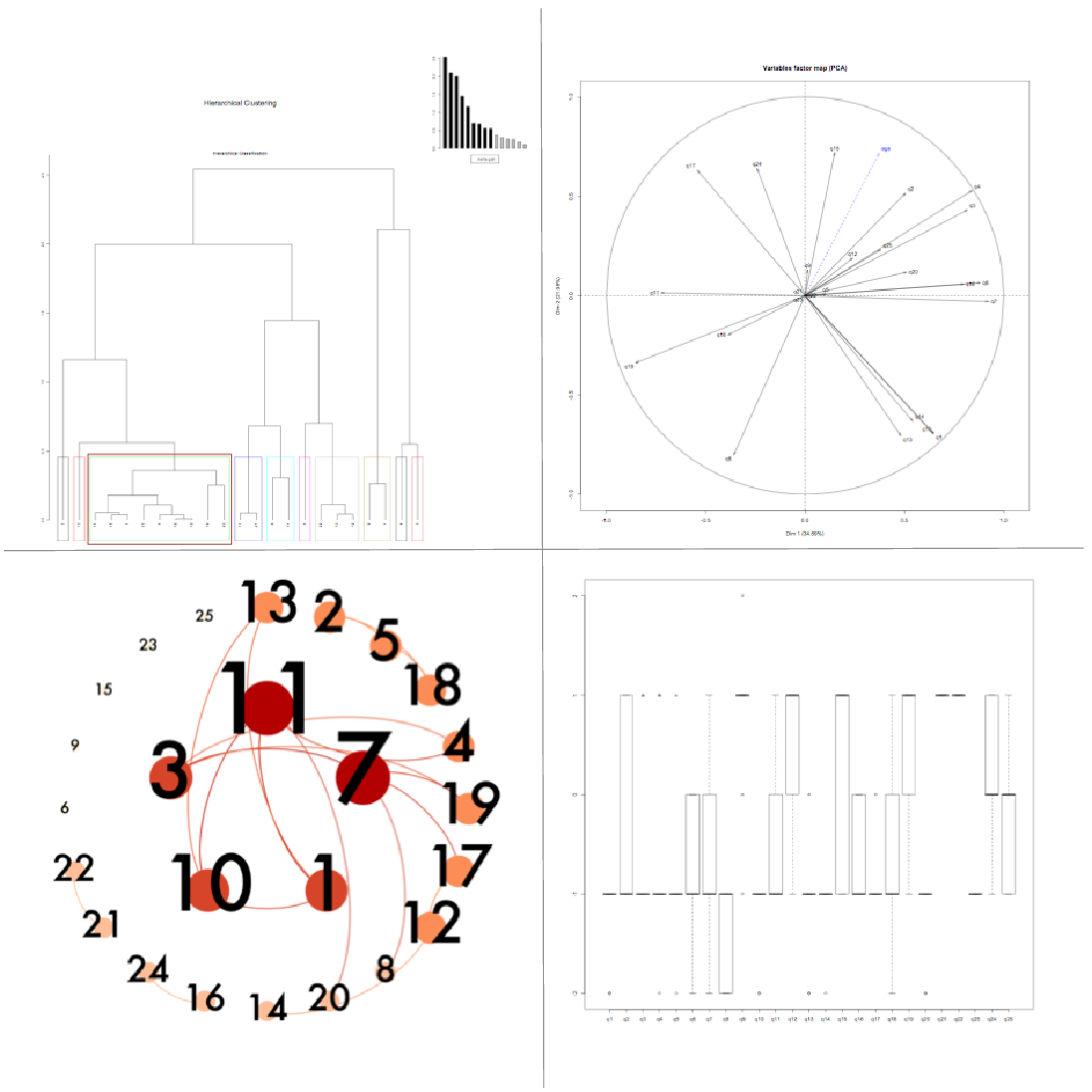}
\caption[]{Examples of results for a coherent set of answers.

 \smallskip

Hierarchical Clustering : The upper left part of the figure shows the result of a hierarchical clustering in one of our sets of answers. Hierarchical clustering is based on the euclidean distance between indviduals considered as a vector with their answers as coordinates. The highlightened group is the group used for this example. 
\smallskip

Principal Component Analysis : The upper right part of the figure shows the PCA results. PCA allows to describe a set of observed variables with a number of new variables called principal components by projecting them in a factorial plan. Two points which are not in the immediate proximity with the circle are not considered because of the projection. The more variables are near the circle, the more there is a principal component which can explain the distribution of the answers and the more coherent is the group. 
\smallskip

Correlation Graph : The lower left part of the figure shows the correlation graph. This graph is builded upon the correlation matrix of all the answers. This graph allows us to determine in a set of answers which one are the most influential in the sense that modifying one of them is susceptible to modify the entire set. The five more influential answers are displayed in the inner circle. 
\smallskip

Boxplots : The lower right part of the figure shows the boxplots for a set of answers. As we are in the case of a coherent set, we can see that for some answers there is a consensus. In complement to the correlation graph, these boxplots allow us to qualitatively determine what are the most important answers for the considered group.}
\label{examples}
\end{figure}

\subsection{Results}
\subsubsection{i. }Statistical analysis (see figure \ref{results}) show that 25\% of the subjects can not be attached to a canonical tradition. This is explained by the fact that these subjects have not provided sufficiently structurated sets of answers or have provided idiosyncrasical structured sets of answers. This result was fully expected because of the nature of our population, the reader has to remember here that the participants were asked to quantify their accordance with a set of propositions which are not from thier field of expertise. This is coherent with the fact that some of them have no structurated approach about our domain. The counterpart of this is the fact that 75\% of our population can be attached to a canonical tradition.

\subsubsection{ii. }Clustering (see figure \ref{examples}) shows that (h0) is verified in so far as, when dividing the general population, the level of structurality tends to increase significatively.

\subsubsection{iii. }
\begin{figure}
\includegraphics{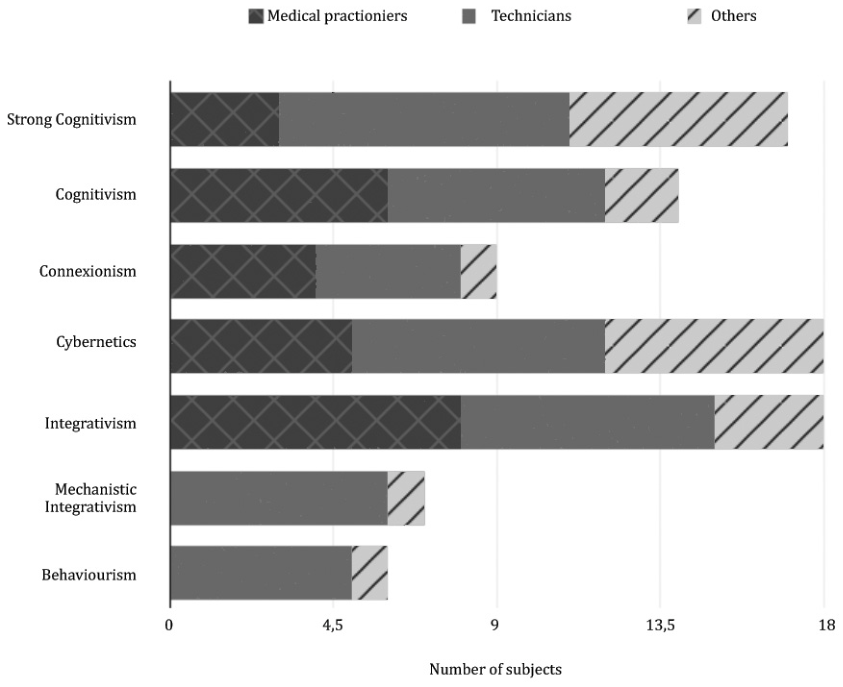}
\caption[]{Canonical traditions repartition

\smallskip

This figure displays the final results of the survey. It gives us a qualitative insight of the clustering of the general population. Subjects non related to a canonical tradition (25\% of the answers) are not represented here because they do not form a coherent group. Is is interesting to note that the ``others'' category is never a majority of a tradition which leads us to conclude that there is not some tradition related to non expertise of the subject.
It is also important to note that there are no medical practitioners in the mechanistic integrativism and in the behaviourism which are two traditions who doesn't have a biological dimension.}\label{results}

\end{figure}

Figure \ref{results} shows that seven different groups can be found inside our general population. Thus we can validate (h1). It is interesting to note that there is no dominant tradition.

\subsubsection{iv. }Appart for two canonical traditions (behaviourism and mechanistic integrativism), the three categories are represented in each tradition. Moreover, medical practitioners are absent from behaviourism and mechanistic integrativism which are two traditions which are deconnected from biological considerations. No professional category has the exclusivity of a tradition or is secluded in one of them. We can then validate (h2).

\subsubsection{v. }We have related some subjects with integrativism by the way of our grid of canonical traditions. Medical practice contains already an integrative dimension in its modes of thinking thus permitting us to validate (h3) because integrativist tradition is not that of only medical practitioners.

\section{Conclusions}

There is not a unique way to conceive what is a human machine system, even inside a given domain of speciality. This also means that, when designing human machine systems, important choices must be made concerning which conceptual framework are used because they can't be interchanged and are not necessary intercompatible.

A grounding framework is meaningful and influences the professional practice of the subject without being a part of it. It is based upon the nature of the objects of the subject's practice. We call that set of ideas \emph{ad hoc conceptual considerations}. 

\emph{Ad hoc conceptual considerations} are an important part of science and engineering because they participate to the practice from a more or less conscious position and are then important to consider in so far as their possible lack of consistancy can potentially lead to serious safety issues. 

Although the usual traditions are present inside the catalog of \emph{ad hoc conceptual considerations}, a significant part of the population is relatable to the integrativist tendancy which is not already structured around a theory. As we have shown in \cite{nazin2014}, integrativism is presently more a way to look at things than a structurated scientifical doctrine. It is nonetheless a good starting point to build a comprehensive and coherent paradigm in human related sciences and engineering taken as a whole field of expertise with its own scientifical principles, methodologies and values. 

\bibliographystyle{plain}
\bibliography{biblio}

\end{document}